\documentclass[a4paper,11pt]{article}
\usepackage{jinstpub} 
\usepackage{lineno}

\title{Performance tests and hardware qualification of the FEBs for the Super-FGD of T2K Phase II}

\author[a,1]{L. Giannessi\note{Corresponding author.} }
\author[a]{, F. Cadoux }
\author[a]{, S. Cap }
\author[d]{, J. Chakrani }
\author[b]{, O. Drapier }
\author[a]{, Y. Favre }
\author[b]{, F. Gastaldi }
\author[c]{, M. Jakkapu }
\author[b]{, J. Nanni }
\author[c]{, K. Sakashita }
\author[a]{and F. Sánchez }

\affiliation[a]{University of Geneva, Section de Physique, DPNC, Geneva, Switzerland}
\affiliation[b]{Ecole Polytechnique, IN2P3-CNRS, Laboratoire Leprince-Ringuet, Palaiseau, France}
\affiliation[c]{High Energy Accelerator Research Organization (KEK), Tsukuba, Japan}
\affiliation[d]{Lawrence Berkeley National Laboratory, US. Before at LLR, France.}

\emailAdd{lorenzo.giannessi@unige.ch}

\abstract{T2K is a long baseline neutrino experiment, entering Phase II with a Near Detector upgrade. The T2K near detector (ND280) upgrade consists of the installation of three new detector systems: a plastic scintillator neutrino active target (Super-FGD), two time projection chambers (HA-TPC) and a time of flight detector (TOF). The Super-FGD is composed of 2-million 1~cm$^{3}$ scintillating cubes read by almost 60 thousand wavelength-shifting (WLS) fibers coupled to an MPPC on one end. Given the large number of channels, the limited space inside magnetic environment, and the limited time from production to installation, the development and testing of the Front-end electronics boards (FEB) for the read-out of the Super-FGD channels represented a challenging task for the success of the upgrade. \\
This work presents the performance tests confirming that the FEB aligns with detector requirements, and the hardware qualification of 240 FEBs through a custom QC test bench designed to detect and locate hardware failures to speed up the repairing process. Installation of the electronics in the detector took place in March 2024, one year after the beginning of the FEB mass production, and the first successful neutrino beam run took place in June of the same year.}


\begin{document}
\maketitle
\flushbottom

\section{Introduction: The Super-FGD}
\label{sec:intro}
T2K~\cite{t2k1,t2k2} is a long-baseline neutrino oscillation experiment that measures muon (anti-) neutrino disappearance and electron (anti-) neutrino appearance from a muon (anti-)neutrino beam. An upgrade of the T2K near detector (ND280), crucial for the reduction of systematic uncertainties in the flux and cross section model, has been completed in summer 2024. Three new subdetectors have been installed, among which the Super-FGD (\textit{Super Fine Grained Detector}) is the active neutrino target. The Super-FGD is a 2-ton fiducial mass scintillating detector composed of 2-million optically isolated polystyrene cubes of 1~cm side. The whole detector is crossed by WLS fibers along the three directions, conveying light to the surface of the detector, where light sensors (SiPMs, or MPPCs) are installed. 


The read-out of the SiPM covering the surface of the detector is handled by 222 Front-End Boards (FEB) organized in 16 crates positioned on the sides of the detector and in a 0.2~T magnetic field. The FEBs amplify, shape and digitize the signals of 55'888 SiPMs, measure signal times, and perform slow control operations. The focus of the present work are the FEB performance tests and hardware qualification of the FEB mass production. 
\subsection{FEB architecture}
\begin{figure}
    \centering
    \includegraphics[width=0.4\linewidth]{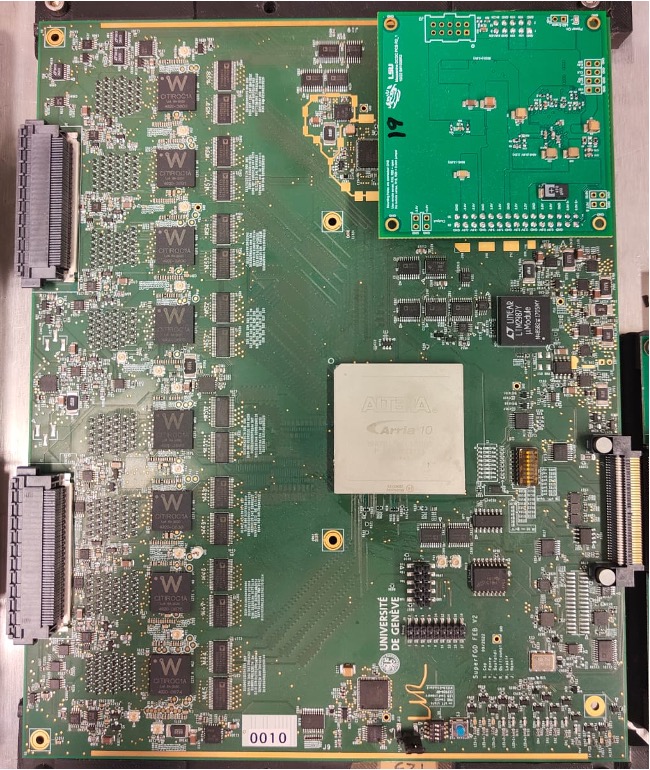}
    \caption{FEB hardware. The eight CITIROCs are visible on the left, while the ARRIA 10 is in the center of the PCB.}
    \label{fig:FEB}
\end{figure}

The FEB employs 8 CITIROCs read-out chips \cite{citiroc2,babymind}, each of which provides timing and charge read-out for 32 channels. Two parallel AC-coupled preamplifiers are employed in the CITIROC (low-gain, LG and high-gain, HG), the signal from the preamplifiers is shaped and the signal amplitude is sampled by a peak detector. Outside of the CITIROC, two 8-ch 12-bit ADCs digitize the CITIROC multiplexed HG and LG output. The CITIROC additionally provides timing information using a constant threshold discriminator on the amplified signal, shaped with a 12.5~ns peaking time shaper. 
DAQ and CITIROC configuration are handled by an FPGA ARRIA 10 by Intel. This also embeds a 400 MHz sampling rate TDC for timing measurement and reads the ADC LG and HG outputs as well as the housekeeping sensors (voltage, current and temperature). The FEB mounts several other active and passive components such as level translators and DC-DC converters for powering, making the FEB PCB relatively dense. A picture of an FEB is shown in figure~\ref{fig:FEB}.

\section{FEB performance tests}
Before starting mass production of PCBs and assembly of components on the FEBs, some tests were carried out to verify that the performance of the FEB are compliant with the detector requirements.
\subsection{Dynamic range}
The Super-FGD must be able to read-out a vast range of signal amplitudes, from the low amplitude LED signals used for calibration, up to the largest proton Bragg peak energy deposits. The required dynamic range is between 1 photoelectron ($\sim$100 fC) and 1000 photoelectron ($\sim$100 pC). 
The lower bound is determined by the trigger threshold, which we are not treating in this study, while the upper bound is due to the saturation of the CITIROC preamplifier, which causes a non-linearity in the CITIROC response. The dynamic range upper limit of the FEB is studied by injecting a signal of known charge in each channel of the FEB, and reading the digitized low-gain CITIROC output for several gain settings.\\ 
The linearity of the output is assessed by requiring a maximum residual of 1\% of the full ADC range with respect to the best fit line describing the data.\\
The results of this study are shown in figure~\ref{fig:dynamic_range}. With a suitable choice of the LG gain, the linear range of the CITIROC response extends well above the required 1000 phototelectrons, satisfying detector requirements.

\subsection{Energy resolution}
\begin{figure}
\begin{minipage}[c]{0.475\linewidth}
    \centering
    \includegraphics[width=\linewidth]{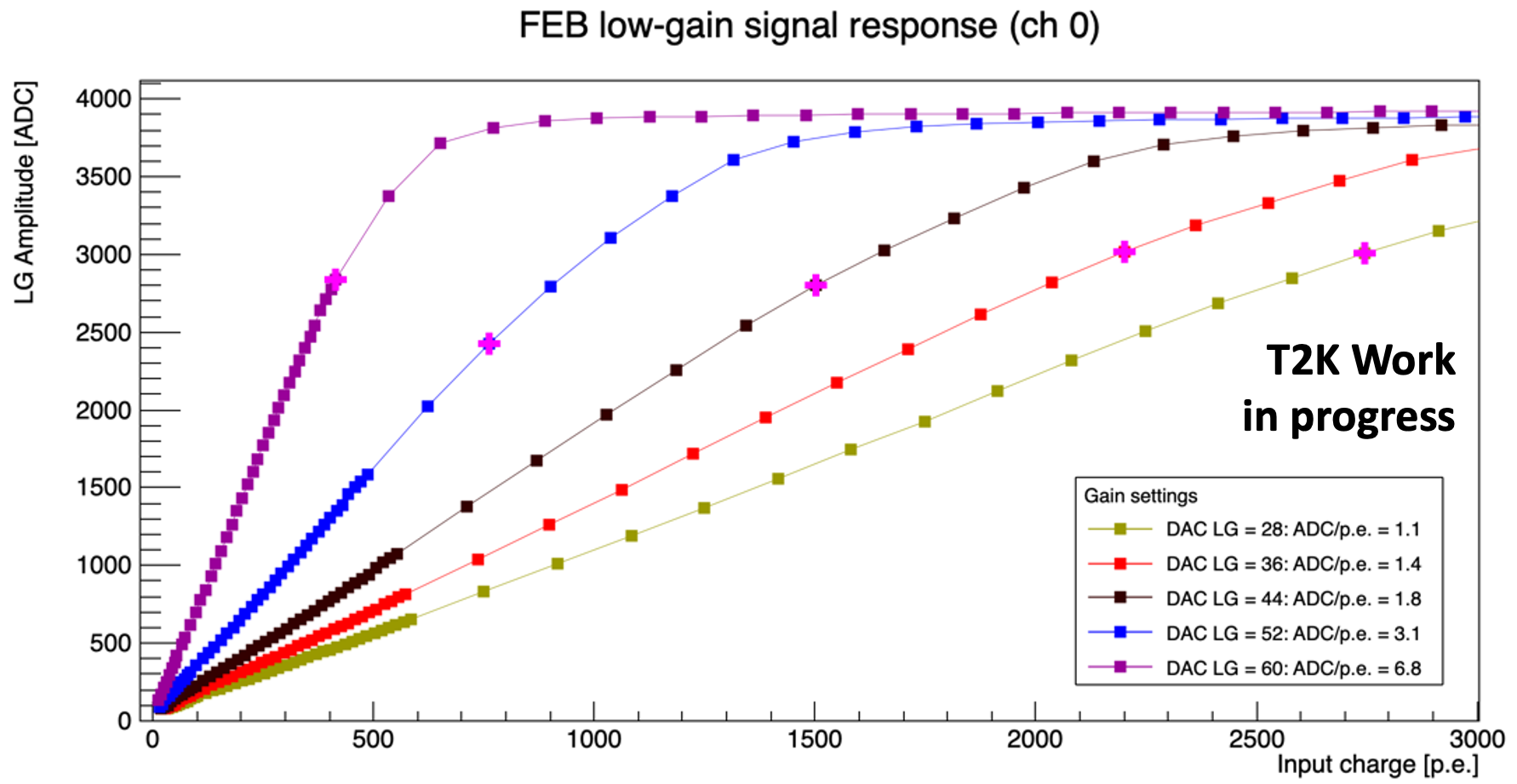}
    \caption{Average of the digitized LG CITIROC response for several injected charge values, and for different gain settings. The ADC/p.e. gain is extracted with Super-FGD calibration using LED light.}
    \label{fig:dynamic_range}
\end{minipage}
\hfill
\begin{minipage}[c]{0.475\linewidth}
    \centering
    \includegraphics[width=\linewidth]{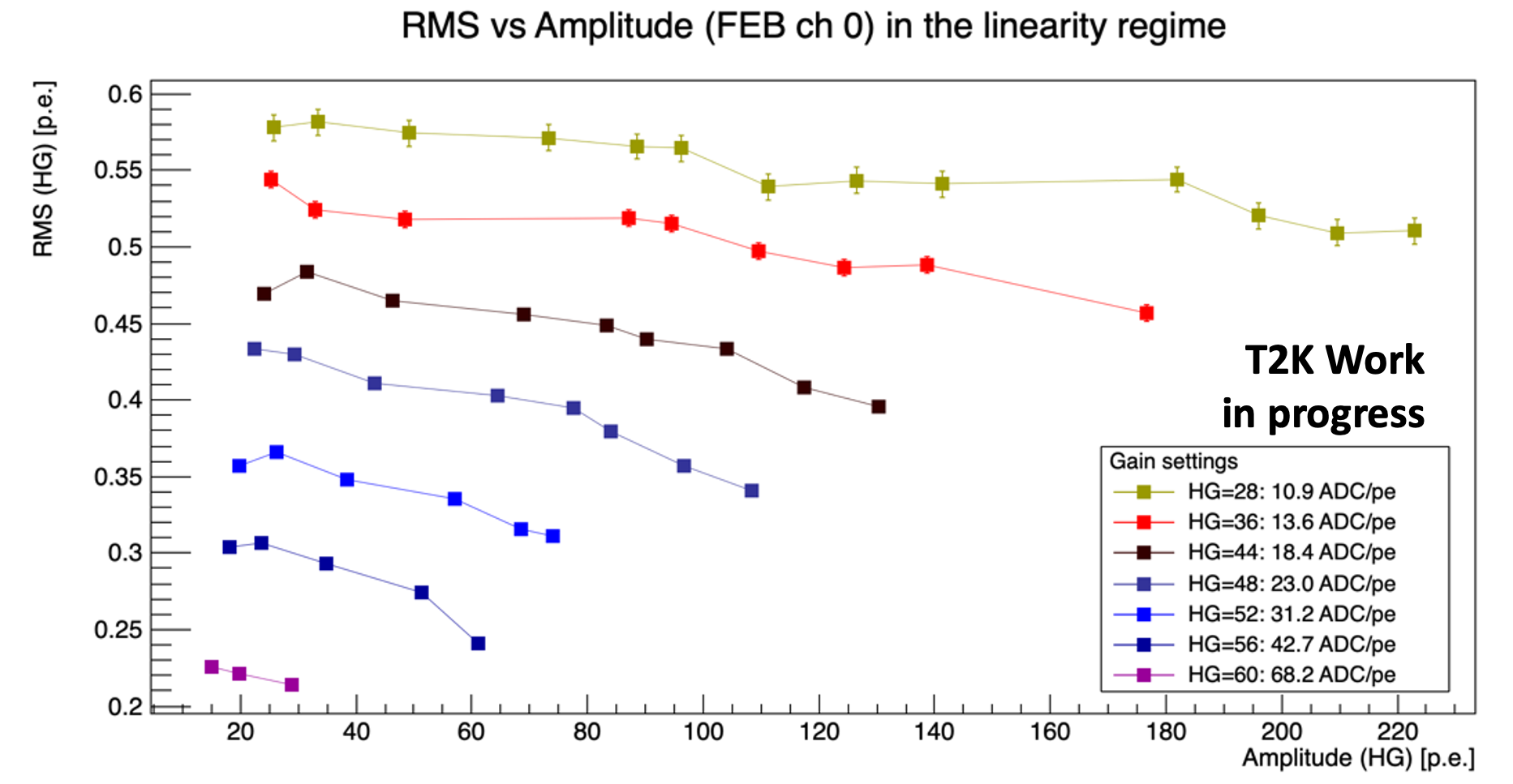}
    \caption{RMS of the digitized HG CITIROC response for several injected charge values and for different gain settings in the linear response regime.}
    \label{fig:energy_resolution}
\end{minipage}
\end{figure}

A similar setup can be used to study the contribution to the energy resolution in the low energy deposit regime (high-gain output), mainly coming from the electronics noise RMS. Injecting a signal of known charge in the linear response regime we can evaluate the standard deviation of the high-gain CITIROC response distribution. The standard deviation of the HG response is below 1 photoelectron for all the gain values used in this study, confirming that electronics noise complies with energy resolution requirements (figure~\ref{fig:energy_resolution}).

\subsection{Electronics cross-talk}
The dominant cross talk between adjacent channels in the Super-FGD is due to the scintillating light leaking between adjacent cubes (optical cross talk). The requirement on the electronics cross talk is to be negligible with respect to the irreducible optical cross talk, expected to be around 3\%~\cite{sfgd_prototype}. The electronics cross talk is studied by injecting a large signal (around 1250 photoelectrons) in one channel while recording the digitized response of all the 256 channels in the FEB. Results are shown in figure~\ref{fig:crosstalk} and show that electronics cross-talk is present exclusively within the same CITIROC, and well below 0.5\% of the injected signal, satisfying design requirements. We believe that the main causes of cross talk are the vicinity of PCB traces in the MIB and internal CITIROC traces. 

\subsection{Timing resolution}
\begin{figure}
\begin{minipage}[c]{0.475\linewidth}
    \centering
    \includegraphics[width=\linewidth]{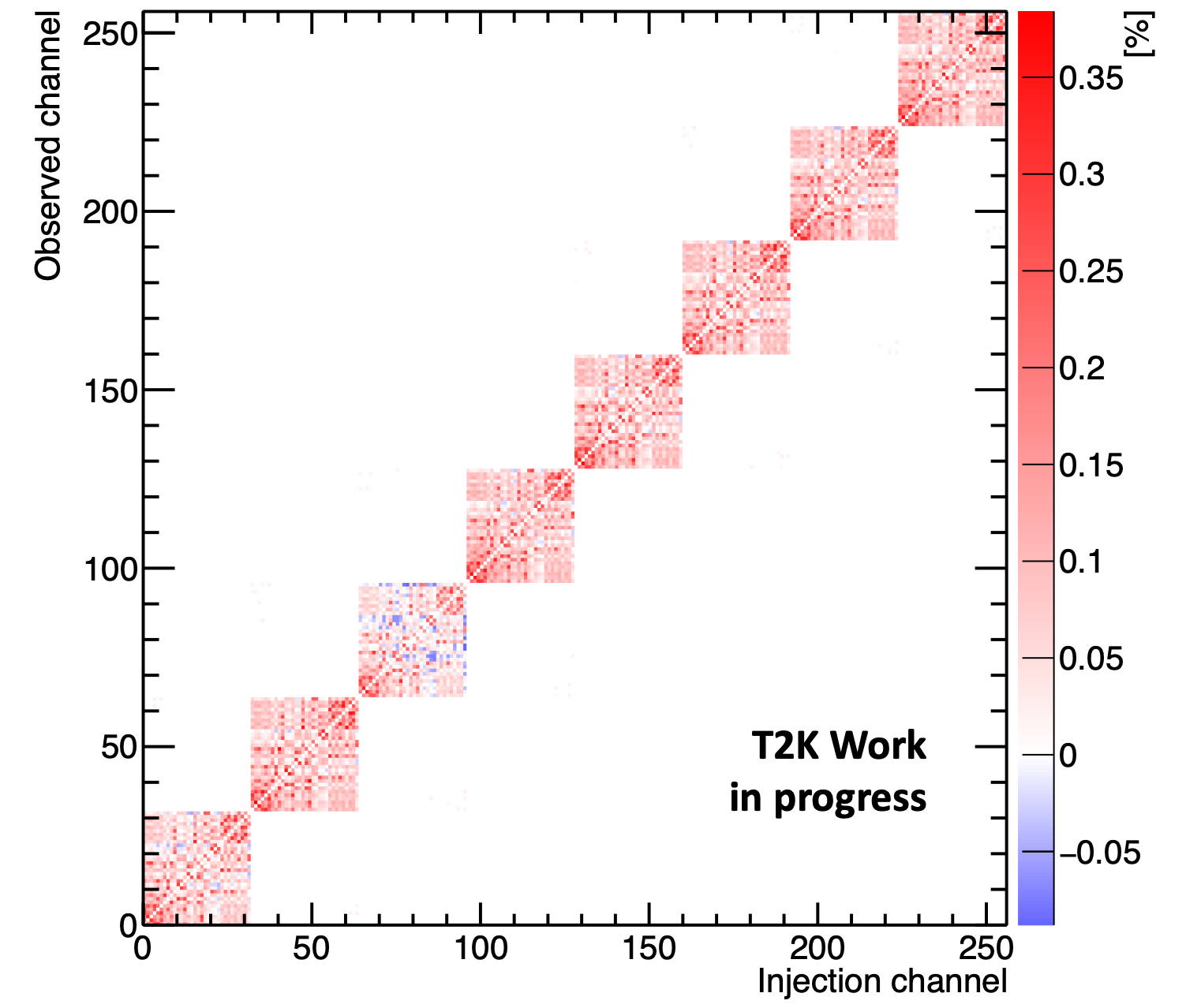}
    \caption{Response of each FEB channel upon injection of a large signal in another FEB channel. Inter-ROC cross-talk effects are negligible.}
    \label{fig:crosstalk}
\end{minipage}
\hfill
\begin{minipage}[c]{0.475\linewidth}
    \centering
    \includegraphics[width=\linewidth]{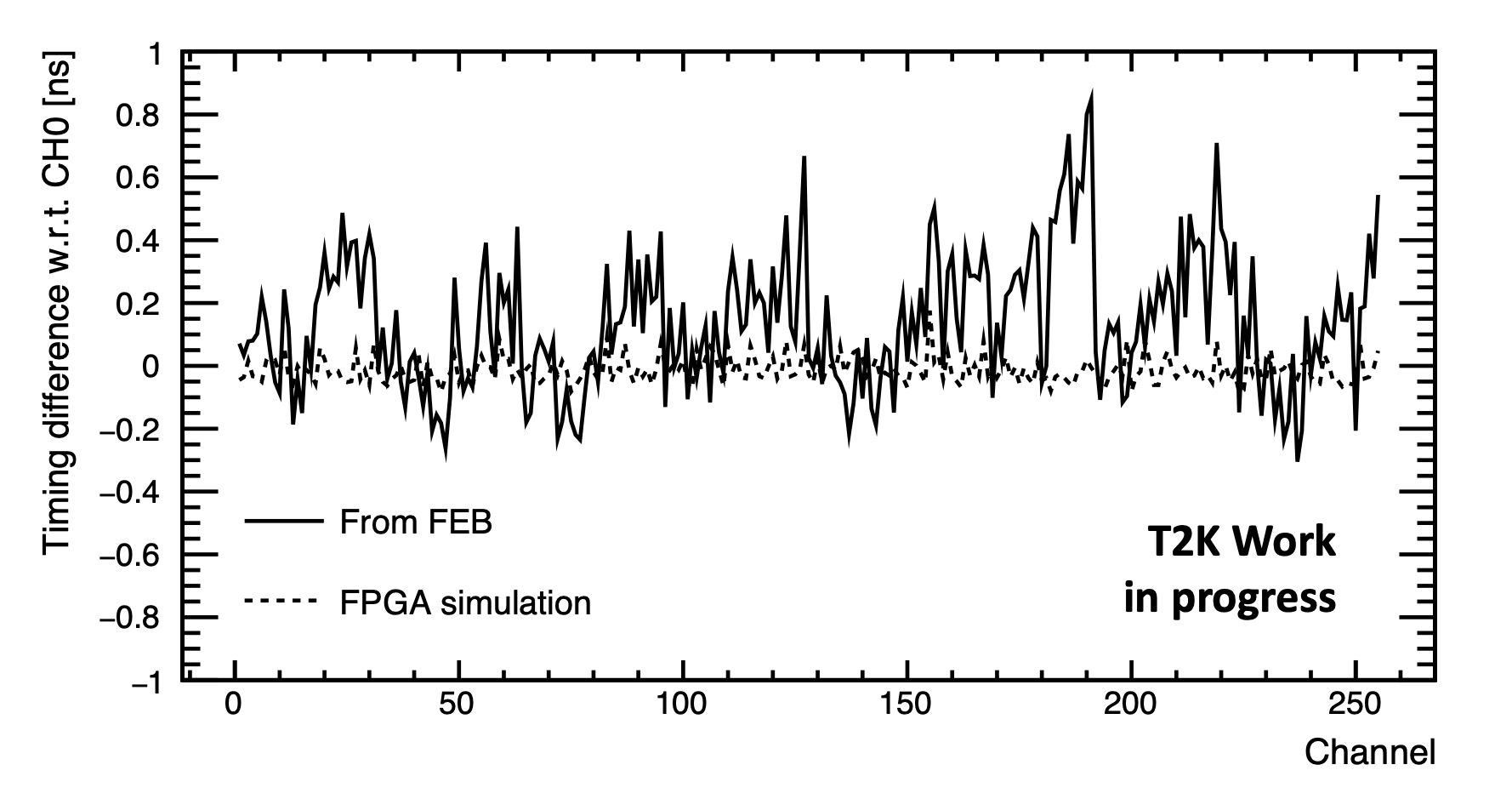}
    \caption{Average time difference between each channel and reference (channel 0) upon injection of synchronised signals. }
    \label{fig:time}
\end{minipage}
\end{figure}

The FPGA employs a 400 MHz sampling rate TDC for signal rising falling time measurements. This imposes a single channel lower limit of $2.5~ns/\sqrt{12}=0.72~ns$ on the time resolution, due to digitization error. However, any time delay between pairs of channels in the detector potentially deteriorates the time resolution. We studied the reciprocal time delay between channels in the FEB by injecting synchronised signals in two FEB channels and measuring the difference between the recorded time stamps. Channel 0 is used as reference. We observe time delays of the order of 1~ns (figure~\ref{fig:time}), potentially deteriorating the time resolution of the system. We believe that these delays are caused by channel-to-channel skew generated in the ASICs and in the level translator chips. It is important to notice that this time delay is a fixed quantity for a given channel, it can thus be calibrated in the full detector system, recovering the nominal time resolution. We conclude that this result, although interesting for the characterization of the FEB and CITIROC, does not represent a deteriorating factor to the time resolution of the Super-FGD.

\section{Mass production hardware qualification}
After carrying out the necessary design tests, such as those outlined in the previous section, green light was given to FEB mass production, which began in March 2023. A total of 240 FEBs were assembled by the beginning of the Summer.\\
The necessity of quickly and reliably testing the hardware of each FEB called for the development of a custom Quality Check (QC) test bench, to test the FEB's hardware and identify possible hardware problems. The goal of the QC test is to check 100\% of the FEB's functionalities and provide information about failure modes, in order to ease the repairing of hardware problems.
\paragraph{QC test bench hardware}
The test bench is a table-top setup composed of several custom boards and devices. The Kaladin\footnote{Developed by SOKENDAI} board handles distribution of the pulse generator signal to each of the 256 FEB channels. The GPIO board (General Purpose Input-Output) controls configuration of the Kaladin board and FEB. It distributes a periodic signal to the FEB for time stamping and transfers the data from the FEB to the PC via USB link. Additionally, there are two adapter boards connecting the GPIO to the FEB, and a power protection board safely sequencing the enabling of FEB, GPIO and Kaladin.
A pulse generator for signal injection, and low and high voltage power supplies complete the setup; they are controlled by the PC via USB link, enabling automatic configuration by the software. 


\paragraph{Functionality test}
The full functionality test is outlined in the following steps.\\
It begins with testing the HV current limiter circuit, halting the the process in cae of failure, to prevent device damage. Next, the connectivity of all digital lines is verified using loop-back circuits. Then, each of the 256 channels is tested against open or short circuits, also confirming that the ADC response aligns with the pulse generator’s amplitude settings. Following this, the functionality of all CITIROC triggers is assessed by observing the ASIC’s behavior across various trigger modes. The final stage involves calibrating the voltage and temperature sensors, ensuring that the standard deviation remains below 0.1 °C for temperature and 0.05 mV for voltage. The functionality test lasts 6 minutes for one FEB.


\subsection{Examples of identified hardware problems}
\begin{figure}
    \centering
    \includegraphics[width=0.8\linewidth]{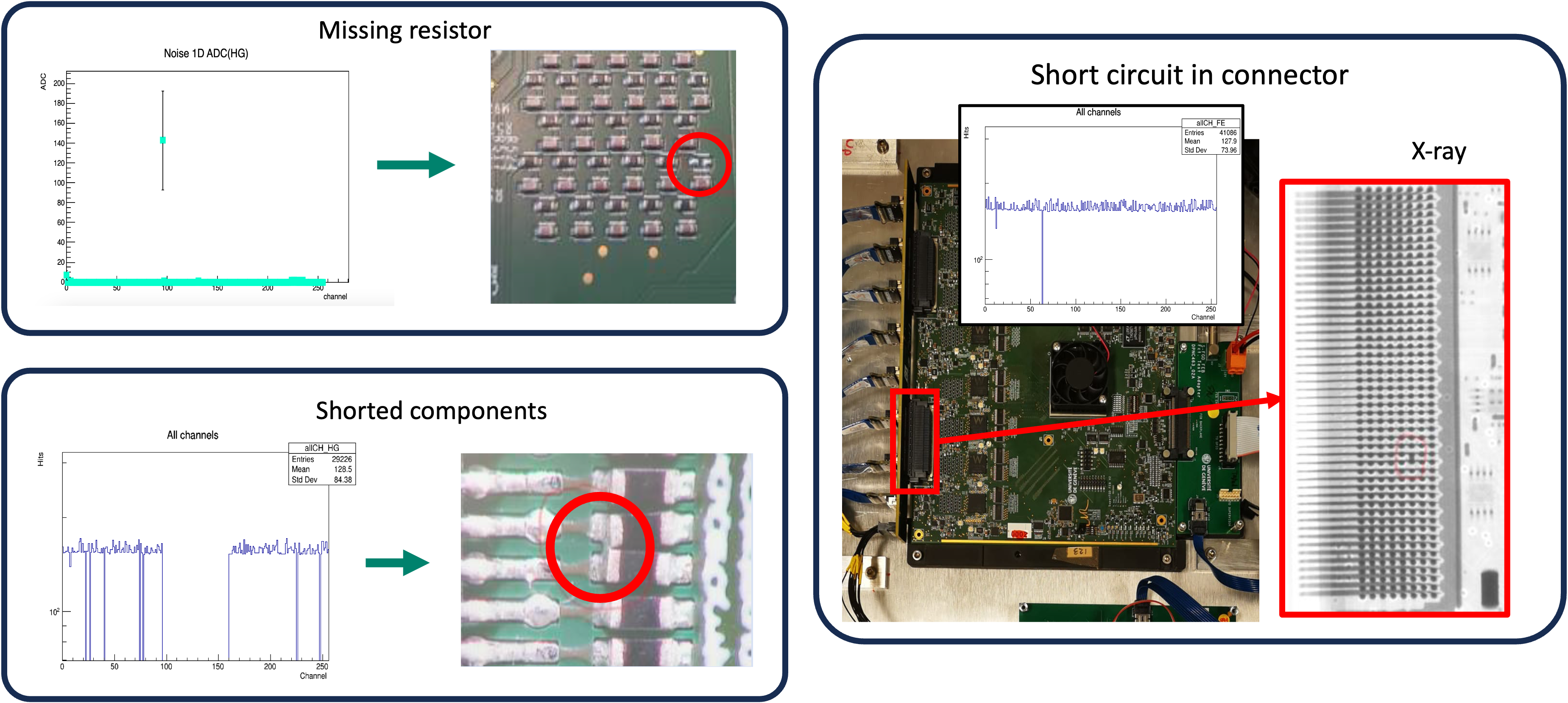}
    \caption{Top left: the abnormally large standard deviation of one channel hints to a missing resistor on the channel read-out chain. Bottom left: the lack of signals in all the channels of 2 adjacent CITIROCs hints to a short between two passive components on the slow control lines handling configuration of those CITIROCs. Right: Lack of signal in one channel caused by a short circuit in the BGA of the MIB connector, between the corresponding channel pin and a ground pin.}
    \label{fig:example1}
\end{figure}

The QC test bench was proved to be particularly useful in locating hardware problems, thus simplifying the repairing process. Some examples are shown in figure~\ref{fig:example1}. In some cases, like those shown on the left-hand side of figure~\ref{fig:example1}, the fix can easily be carried out in a small workshop, effectively speeding up the repairing process. In other cases, like the one shown on the right-hand side of figure~\ref{fig:example1}, the QC test helps finding critical processes in the assembly procedure to be communicated to the manufacturer.

\section{Conclusions}
The read-out electronics system of the Super-FGD posed the challenge of developing a dense Front-End Board, and validating the hardware of 240 FEBs in a relatively short amount of time. Performance tests were successfully carried out, verifying the electronics design, and then a custom QC test bench was developed for testing FEBs and locating hardware problems. After testing and repairing of the damaged boards hardware, the 222 FEBs necessary for the detector read-out (along with 18 additional spares) were installed in March 2024. The first successful neutrino beam run of T2K Phase II followed in June 2024.





\bibliographystyle{JHEP}
\bibliography{main.bib}

\providecommand{\href}[2]{#2}\begingroup\raggedright\begin{thebibliography}{1}

\bibitem{t2k1}
{Abe, K.}, {Akhlaq, N.}, {Akutsu, R.}, {Ali, A.}, {Alonso Monsalve, S.}, {Alt, C.} et~al., \emph{Measurements of neutrino oscillation parameters from the t2k experiment using $3.6×10^{21}$ protons on target}, \href{https://doi.org/10.1140/epjc/s10052-023-11819-x}{\emph{Eur. Phys. J. C} {\bfseries 83} (2023) 782}.

\bibitem{t2k2}
K.~Abe, N.~Abgrall, H.~Aihara, Y.~Ajima, J.~Albert, D.~Allan et~al., \emph{The t2k experiment}, \href{https://doi.org/https://doi.org/10.1016/j.nima.2011.06.067}{\emph{Nuclear Instruments and Methods in Physics Research Section A: Accelerators, Spectrometers, Detectors and Associated Equipment} {\bfseries 659} (2011) 106}.

\bibitem{citiroc2}
D.~Impiombato, S.~Giarrusso, T.~Mineo, O.~Catalano, C.~Gargano, G.~{La Rosa} et~al., \emph{Characterization and performance of the asic (citiroc) front-end of the astri camera}, \href{https://doi.org/https://doi.org/10.1016/j.nima.2015.05.028}{\emph{Nuclear Instruments and Methods in Physics Research Section A: Accelerators, Spectrometers, Detectors and Associated Equipment} {\bfseries 794} (2015) 185}.

\bibitem{babymind}
O.~Basille et~al., \emph{{Baby MIND Readout Electronics Architecture for Accelerator Neutrino Particle Physics Detectors Employing Silicon Photomultipliers}}, \href{https://doi.org/10.7566/JPSCP.27.011011}{\emph{JPS Conf. Proc.} {\bfseries 27} (2019) 011011}.

\bibitem{sfgd_prototype}
A.~Blondel, M.~Bogomilov, S.~Bordoni, F.~Cadoux, D.~Douqa, K.~Dugas et~al., \emph{{The SuperFGD Prototype charged particle beam tests}}, \href{https://doi.org/10.1088/1748-0221/15/12/P12003}{\emph{Journal of Instrumentation} {\bfseries 15} (2020) P12003}.

\end{thebibliography}\endgroup


\end{document}